# *Expansion of the Kullback-Leibler Divergence,*

# *and a new class of information metrics*

David J. Galas[1*], T. Gregory Dewey[2], James Kunert-Graf[1] and Nikita A. Sakhanenko[1]

[1]Pacific Northwest Research Institute
720 Broadway
Seattle, Washington 98122

[2]Albany College of Pharmacy and Health Sciences
106 New Scotland Avenue
Albany, New York 12208

* *Communications to: dgalas@pnri.org*

**Abstract:** Inferring and comparing complex, multivariable probability density functions is fundamental to problems in several fields, including probabilistic learning, network theory, and data analysis. Classification and prediction are the two faces of this class of problem. We take an approach here that simplifies many aspects of these problems by presenting a structured, series expansion of the Kullback-Leibler divergence - a function central to information theory - and devise a distance metric based on this divergence. Using the Möbius inversion duality between multivariable entropies and multivariable interaction information, we express the divergence as an additive series in the number of interacting variables, which provides a restricted and simplified set of distributions to use as approximation and with which to model data. Truncations of this series yield approximations based on the number of interacting variables. The first few terms of the expansion-truncation are illustrated and shown to lead naturally to familiar approximations, including the well-known Kirkwood superposition approximation. Truncation can also induce a simple relation between the multi-information and the interaction information. A measure of distance between distributions, based on Kullback-Leibler divergence, is then described and shown to be a true metric if properly restricted. The expansion is shown to generate a hierarchy of metrics and connects this work to information geometry formalisms. We give an example of the application of these metrics to a graph comparision problem that shows that the formalism can be applied to a wide range of network problems, provides a general approach for systematic approximations in numbers of interactions or connections, and a related quantitative metric.

________________________________________________________________________________

**Introduction**

The problem of representing or inferring dependencies among variables is central to many fields. It is fundamental to data analysis of large data sets, as well as describing and approximating the behavior of physical, chemical, and biological systems with many modes, particles, or component interactions. These complex systems are usually modeled by graphs or hypergraphs, and their inference from data represents a central problem. Statistical inference and machine learning approaches have been directed at this general class of inference problem for many years, and the literature of physical chemistry, among other, related fields, abounds with approaches to the general representation problem [6,7,11,12]. A key related problem is that of measuring the difference between approximations, a useful metric of probability distributions. The relationships between neural networks, statistical mechanics, and this general class of problem have also been explored [13]. While certainly not the only indication of complexity, the number of variables that interact or are functionally interdependent, is a very important characteristic of the complexity of a system. We engage a number of these problems in this work.

A central function of information theory, the Kullback-Leibler divergence, can be shown to be close to the heart of these problems. It is the goal of this paper to describe an approach that simplifies some aspects of these problems in a different way, by focusing on interesting and useful symmetries of entropy and "relative entropy" and the Kullback-Leibler divergence (K-L, referred to as the "divergence" in the rest of this paper.) Particularly important in practical applications is the divergence between the "true", multivariable, probability density function (pdf) and any approximation of it [1], as are specific metric measures of the distances between approximations.

The paper is structured around recognition and exploitation of several properties of this divergence, a central function in information theory. We first show that the divergence admits of a simple series expansion with increasing numbers of variables in each successive term. We affect this expansion of the multivariable cross-entropy (or relative entropy) term of the divergence using the Möbius duality between multivariable entropies and multivariable interaction information [2-5]. This allows a series expansion in the number of interacting variables, which can be used as an approximation parameter: the more interactions considered, the more accurate the approximation. We then illustrate the derivation of some known factorizations of the pdf by truncating the expansion at small numbers of variables. Well-known simple approximations emerge, including the Kirkwood superposition approximation at three variables. This is a widely used approximation in the theory of liquids [6-8]. Other approximations, like the seminal approximation method of Chow and Liu [15], is closely connected to the expansion. We will not expand on this specific approach here, but will explore and extend this connection in future work.

The divergence expansion we propose is entirely general, can be extended to any degree, and leads to a number of useful relationships with other information theory measures. In the following section we define a new simple metric between probability density functions and show that it meets all the requirements of a true metric.

Unlike the approach of the Jensen-Shannon divergence - which is a measure based on symmetrizing the K-L divergence [16] - or that which use the Fisher metric to embed the functions in a Riemannian manifold [14], our metric provides a large class of information metrics that calculate distances directly, and thereby easily measure the relations between approximations, among other applications. We examine a few cases of specific pdf function classes (e.g., Gaussian, Poisson) and find explicit forms for the functions. Finally, we examine briefly the metric distances implied by different truncations of the divergence expansion, and describe an application to the character description of networks.

**Expanding the divergence.**

Consider a set of variables $v = \{X_i\}$, for which we have many values constituting a data set. The concepts of maximum entropy and minimum divergence have been used to devise approaches to the inference of the best estimate of the true probability density function from a data set. The relation between the "true" and an approximate probability density function (pdf) is best characterized by the Kullback-Leibler divergence. If the true pdf is $P(v)$ and an approximation to it is $P'(v)$ then the divergence is given by

$$D(P \parallel P') = \sum_s P(s) \log\left(\frac{P(s)}{P'(s)}\right),$$

(1)

where $s$ traverses all possible states of $v$. The approximated entropy (called the cross-entropy) is defined as

$$H'(v) = -\sum_s P(s) \log(P'(s)),$$

(2)

so the divergence is simply the difference between the true entropy and the cross entropy:

$$D(P \parallel P') = H'(v) - H(v) \qquad (3)$$

In this form it is clear that the approximate joint entropy must be greater than H(v) since we know the divergence is always non-negative [1]. This is a consequence of the well-known Jensen's inequality. If *P'* is an approximation to *P*, then as the approximation gets better and better the divergence converges to zero. The approximation of the joint entropy is the measure of the accuracy of the approximation and minimizing *H'* (under some set of constraints or assumptions) must be optimum. Using other information theory measures related to the joint entropies in Equation 3, however, can also be used to good effect.

Specifically, we use the Möbius inversion relation between the entropy and interaction information [3-5]. This relationship can be written

$$H(v) = \sum_{\tau \subseteq v} (-1)^{|\tau|+1} I(\tau), \qquad (4a)$$

where the sum is over all subsets of *v*. H and *I* can be exchanged in this symmetric form of the relation and the equation still holds.

$$I(v) = \sum_{\tau \subseteq v} (-1)^{|\tau|+1} H(\tau) \qquad (4b)$$

The symmetry derives from the inherent structure of the subset lattice, which is a hypercube [9]. Inserting the joint entropy expression into Equation 3 gives a sum over all subsets of the variables

$$D(P \parallel P') = \sum_{\tau \subset v} (-1)^{|\tau|+1} I'(\tau) - H(v) \qquad (5a)$$

Now if we group terms by the number of variables in the subset and introduce notation to indicate the size of each of the subsets, the sum is rearranged as an expansion.

$$D(P \parallel P') = \sum_{m=0}^{|v|} \sum_{\tau_m \subset v} (-1)^{|\tau_m|+1} I'(\tau_m) - H(v)$$

(5b)

The symbol $\tau_m$ indicates a subset of variables of cardinality $m$ ($|\tau_m|=m$). This then becomes an expansion in degrees, $m$, the number of variables. The full expansion includes, and terminates with, the full set of variables, $v$.

**Truncations of the series.**

If we truncate the expansion at various degrees (numbers of variables), setting all interaction information terms above the truncation point equal to zero, we generate a series of increasingly accurate, but ever more complex, approximations. Truncation generates a specific probability density function relation, in the form of a specific factorization, by setting to zero an interaction information expression in the form of a sum of entropies.[1] This is a key result of the expansion of the divergence. Truncation, and a factorization of the probability density function, results from setting all the higher interaction informations to zero. Thus, the expansion represents a method for approximation and simplification that specifically limits the degree of variable dependencies.

The approximations that result from truncating the expansion of the cross entropy at the first few degrees are familiar ones. Since the number of dependent variables is the driver of the complexity, we begin with pairwise approximations and stepwise increase the number. The first few truncations show the character of this expansion process.

Truncation at *m=1*

Considering the simplest possible truncation, setting all but the first term equal to zero:

$$D(P \parallel P') = D_1(P \parallel P') = \sum_i H'(X_i) - H(v) \equiv A_1 - H(v). \tag{6a}$$

---

[1] Note that setting $I(\tau_m)=0$ does not imply that higher terms, $I(\tau_{m+1})$ etc., are also zero. The truncation approximation necessarily sets all higher terms to zero.

This truncation requires that all of the *m*=2 terms, for all pairs, are zero

$$I'(X_i, X_j) = 0 \quad \forall i,j \tag{6b}$$

which from equations 2 and 4b implies independence of all pairs of variables, and the simplest factorization:

$$P(X_i, X_j) = P(X_i)P(X_j), \quad \forall X_i, X_j \in \nu \tag{6c}$$

This determines all pairwise probability functions, but note that it actually does not determine the form of a three-way or higher pdf. The truncation requires, however, that the three-variable interaction information is zero: $I(X_i, X_j, X_k) = 0$. This fact combined with Equations 4b and 6b results in a full three-way factorization of the pdf:

$$P(X_i, X_j, X_k) = P(X_i)P(X_j)P(X_k). \tag{7a}$$

As variables are added we can use the interaction information recursion relation (Equation 12) to derive the higher pdf's implied by the truncation. Finally, the *m*=1 truncation yields the fully factored pdf

$$P(\nu) = \prod_i P(X_i) \tag{7b}$$

Note that this same result derives from minimizing the expression for the divergence in 6a, since this expression is a minimum when $\sum_i H(X_i) = H(\nu)$.

The implication of this result for data analysis is simply the solution of the classic problem of determining the optimal pdf under the assumption of the independence of all variables, fixed expectation values being defined by parameters usually represented by Lagrange multipliers. The physics implication would be simply that of independent particles, observables, etc., which leads to a simple Boltzmann distribution in equilibrium. The pdf becomes more complex, of course, if we truncate the expansion at a higher level.

Truncation at *m*=2

This truncation requires that $I'(X_i, X_j, X_k) = 0$, which from Equation 4b implies this factorization

$$P_2'(\tau_2) = \prod_{i>j} \frac{P(X_i, X_j)}{P(X_i)} = \prod_{i>j} P(X_i | X_j) \quad \forall \tau_2 \subset v \tag{8a}$$

Let us denote the cross entropy term for this truncation as $A_2$. Then we have

$$D_2(P \parallel P') = \sum_i H'(X_i) - \sum_{\tau_2 \subseteq v} I'(\tau_2) - H(v) \equiv A_2 - H(v) \tag{8b}$$

$$A_2 = -\sum_{i>j}\left(H'(X_i) + H'(X_j) - H'(X_i, X_j)\right) + \sum_i H'(X_i) \tag{8c}$$

The cross entropy term $A_2$ is determined by the pdf $P'$, and from 8b above we can see that the minimization of the divergence is the same as truncation of the expansion. This is equivalent to the approximation made by Chow-Liu [15]. In physical terms this is the same as ignoring all but pairwise interaction terms in a Hamiltonian, and is precisely the probabilistic version of the Kirkwood superposition approximation [6-8]. This approximation is used in the physics of dense multiparticle systems, like liquids. The resulting pair correlation function is used in deriving many of the thermodynamic properties of liquids. Singer [6] related this to the more general theoretical constructs like the Percus-Yevick approximation and the Bogoliubov-Born-Green-Kirkwood-Yvon (BBGKY) hierarchy.

Truncation at *m=3*

Parallel to the above we can express the truncation approximation at the next level using three terms:

$$D_3(P \parallel P') = \sum_i H'(X_i) - \sum_{\tau_2 \subseteq v} I'(\tau_2) + \sum_{\tau_3 \subseteq v} I'(\tau_3) - H(v) \equiv A_3 - H(v) \tag{9a}$$

In terms of the cross entropies the term $A_3$ becomes

$$A_3 = \sum_{i>j>k} \left(H'(X_i)+H'(X_j)+H'(X_k)-H'(X_iX_j)-H'(X_iX_k)-H'(X_jX_k)+H'(X_iX_jX_k)\right) -$$
$$\sum_{i>j}\left(H'(X_i)+H'(X_j)-H'(X_i,X_j)\right)+\sum_j H'(X_j)$$

(9b)

As before we see that the truncation, assuming the four-variable cross-interaction information is zero, is the same as minimizing the divergence $D_3$. Both imply that the approximation to the pdf is

$$P_3'(v) = \prod_{i>j>k} \frac{P(X_iX_jX_k)P(X_k)P(X_j)}{P(X_iX_k)P(X_jX_k)} \qquad (10)$$

Note that $A_3$ is also expressed simply in terms of the deltas used in the analysis of dependency and as a partial measure of complexity [10]. For three variables this quantity is the same as the conditional mutual information, as can be seen from the recursion relation, equation 12.

$$A_3 - \sum_i H'(X_i) = -\sum_{k;i>j} I(X_iX_j|X_k) \qquad (11)$$

The approximation indicated on the right-hand side is based on the cross entropy approximation, that $H = H'$. The truncation of the expansion, leading to more complex representations of the variable interactions, can be taken to higher levels, of course, which leads in turn to higher-level, more complex, factorizations of the pdf. These factorizations are most simply seen by setting the cross interaction information for $m$ variables equal to zero and inferring the implied pdf factors.

**A relation to the deltas.**

The truncation relation implies another simple equivalence that has direct intuitive meaning, and connects in a simple way to the differential interaction information [10]. From the general recursion relation for the interaction information we can derive a set of simple equivalences.

For the set $v_n$ of $n$ variables the general, multi-variable recursion relation for the interaction information is

$$I(v_n) = I(v_{n-1}) - I(v_{n-1} | X_n) \tag{12}$$

for all $n$ choices of $X_n$, where the set $v_{n-1}$ is the set missing $X_n$. Thus the truncation, setting the left side to zero, implies exactly $n$ relations, one for each choice of $i$:

$$I(v_{n-1} | X_i) = I(v_{n-1}) \tag{13}$$

The implication of the truncation criterion for the divergence at $m=n$, then, is that the interaction information, conditioned on each variable of a set $v_n$, is the same as the interaction information of the remaining $n$-1 variables. Note that the conditional in Equation 12 is the same (within a sign) as the asymmetric delta function for $n$ variables [10], so the truncation of the divergence is seen to be equivalent to a simplification and truncation of the asymmetric delta. For truncation at $m=2$ this would mean that all conditional mutual informations are equal to the mutual information itself: equivalent to specifying independence of the conditional variable.

**Multi-information.**

It is easy to show that the truncation embodied in Equation 11 also implies a simple relation between the "multi-information" (called "complete correlation" by Watanabe [11]) and the interaction information. The multi-information is defined as $\Omega(v_n) = \sum_i H(X_i) - H(v_n)$. This quantity is often used as a measure of overall multivariable dependence, since it goes to zero if all variables are independent. It is always positive, but has several drawbacks in that it does not distinguish at all the degrees of dependence (number of variables), and is not a metric.

We will not show the elementary proof of the general case of truncation at $n$ variables here, but illustrate a simple expression for multi-information in terms of the interaction information with the 3- and 4-variable cases. For the case of truncation at $n=3$ ( $I(v_3) = 0$ ) the relation is simply the sum of all three mutual informations:

$$\Omega(X,Y,Z) = I(X,Y) + I(X,Z) + I(Y,Z) \qquad (14a)$$

This is easy to see by direct calculation using the marginal entropies. For n=4

$$\Omega(X,Y,Z,W) = I(X,Y,Z) + I(X,Y,W) + I(X,Z,W) + I(Y,Z,W) - I(X,Y) - I(X,Z) -$$
$$I(X,W) - I(Y,Z) - I(Y,W) - I(Z,W) = \sum_{\tau_3} I(\tau_3) - \sum_{\tau_2} I(\tau_2) \qquad (14b)$$

The relation 14a is strongly intuitive in the sense that if the 3-variable interaction information is zero, the multi-information is simply the sum of the mutual information for all three pairs. A similar, but less intuitive, relationship is embodied in the 4-variable case, Equation 14b, and the general case is suggested.

The divergence expansion can also be expressed using the multi-information in a limited number of variables, as well as a series of truncation-approximate probability density functions, in the following way. Consider a series of functions $\{P_m\}$ related to the true, untruncated, probability density function, such that $P_m$ is the pdf of $m$ variables that results from setting the interaction information equal to zero for subsets $\tau_m$. Then we have

$$I(\tau_m) = 0 \Rightarrow H(\tau_m) = \sum_{\eta \subset \tau_m}(-1)^{|\eta|+1} H(\eta); \ \tau_m \subseteq \nu, |\nu| = n. \qquad (15)$$

The divergence converges to zero for the series $\{P_m\}$ as the number of variables increases to $n$.

$$\lim_{m \to n} D(P \parallel P_m) = 0 \qquad (16)$$

The divergence therefore induces a topology on the series of functions. The proof of 16 follows directly from the definitions.

Note that the multi-information is not a metric, and that a metric specifically gives a distance measure between different pdfs. This is a problem that has received much attention as a metric provides a clear measure of the function space, we can complete this formalism around the K-L divergence and its approximations by devising a simple pdf metric.

**Information Geometry and a simple metric.**

Although it is sometimes thought of as a distance measure between probability distributions, the Kullback–Leibler divergence is not a true metric. Among the disqualifying properties is its asymmetry. There has, however, been much work devoted to the development of geometric measures of information, particularly in differential geometry [14], and symmetric divergences have been defined [16]. A derivative form, the Hessian, of the divergence does yield a metric tensor known as the Fisher information metric. This is a Riemannian metric tensor, and has been used extensively. While having a real metric is essential to a complete quantitative theory, it is even more useful if it is relatively simple and direct. Finite distances between functions in the differential manifold of the Fisher metric must be determined by integration along geodesics. Simpler metrics allow the direct calculation of the distance between probability density functions. We now describe such a simple information metric.

Consider the problem of comparing two approximate distributions, $R(v)$ and $S(v)$ using another pdf, $P(v)$, as a reference function. We use the K-L divergence to define a metric simply as the absolute value of the difference between two K-L divergences using the same reference function. This definition is embodied in the following equation.

$$\mathfrak{D}_P(R \parallel S) \equiv |D(P \parallel R) - D(P \parallel S)| = \left| \sum_s P(s) \, log\left(\frac{R(s)}{S(s)}\right) \right|$$

(17)

We next establish that $\mathfrak{D}_P(R \parallel S)$ does indeed have the properties of a metric on a function space. A metric has the following four properties, which we show are fulfilled by our definition:

1. Non-negativity: $\mathfrak{D}_P(R \parallel S) \geq 0$ is assured because $P(v) \geq 0$ and the absolute value in Equation 17 assures a summation that is non-negative.

2. Identity of indiscernibles: $\mathfrak{D}_P(R \parallel S) = 0$ when $R(v) = S(v)$. $\mathfrak{D}_P(R \parallel S) = \left| \sum_s P(s) \left( \log\left(\frac{R(s)}{P(s)}\right) - \log\left(\frac{S(s)}{P(s)}\right) \right) \right| = 0.$ For a metric it must also be true that $\mathfrak{D}_P(R \parallel S) \neq 0$, unless $R(v) = S(v)$, otherwise the metric is a pseudometric. This condition does not hold for all choices of P, R and S and therefore the metric property may apply only to specific spaces, and must be examined in each case. We illustrate this later for some specific cases.

3. Symmetry: $\mathfrak{D}_P(R \parallel S) = \mathfrak{D}_P(S \parallel R)$.

$$\mathfrak{D}_P(R \parallel S) = \left| \sum_s P(s) \left\{ \log\left(\frac{R(s)}{P(s)}\right) - \log\left(\frac{S(s)}{P(s)}\right) \right\} \right| = \mathfrak{D}_P(S \parallel R)$$

4. Subadditivity, obeying the triangle inequality:

$$\mathfrak{D}_P(R \parallel S) \leq \mathfrak{D}(R \parallel Q) + \mathfrak{D}(Q \parallel S)$$

$$\mathfrak{D}_P(R \parallel S) = \left| \sum_s P(s) \log\left(\frac{R(s)}{S(s)}\right) \right| = \left| \sum_s P(s) \log\left(\frac{R(s)}{Q(s)}\right)\left(\frac{Q(s)}{S(s)}\right) \right|$$

$$= \left| \sum_s P(s) \log\left(\frac{R(s)}{Q(s)}\right) + \sum_s P(s) \log\left(\frac{Q(s)}{S(s)}\right) \right|$$

$$\leq \left| \sum_s P(s) \log\left(\frac{R(s)}{Q(s)}\right) \right| + \left| \sum_s P(s) \log\left(\frac{Q(s)}{S(s)}\right) \right| = \mathfrak{D}_P(R \parallel Q) + \mathfrak{D}_P(Q \parallel S)$$

The inequality holds because the sums are real numbers, and the triangle inequality applies. $\mathfrak{D}$ is therefore a true metric on the function space of pdf's, which we can use directly as a measure of information distance. In some cases and function spaces, however, there are subspaces that are true metrics and other that are pseudometrics, having some distinct functions that have zero distance from each other. Since the metric is determined by a reference function, $\mathfrak{D}$ represents a large class of metrics, each determined by the choice of reference function. We now examine some properties of these metrics.

An intriguing similarity of the metric, the distance between functions defined by a third function, lies in Bayesian statistics. We could say that by defining the reference pdf, $P(v)$ as a prior pdf, $\mathfrak{D}(R \parallel S)$ measures the distance between two posterior functions, $R(v)$ and $S(v)$. By measuring the distance between successive posteriors, one can monitor the convergence of Bayesian updating to a steady state distribution. The distance measure can also be used to assess quantitatively how close different posterior models are to each other. We could define a Dirichlet metric, $\mathfrak{D}_D$ for example, if the reference pdf, or prior, were a Dirichlet distribution, or a uniform, or a Gaussian metric if the reference were uniform or Gaussian.

**Special Metrics.**

The fact that $P(v)$ defines a metric on a function space inspires us to ask what specific functional forms yield metric spaces with particular properties. We could define a uniform probability density over the variable set $v$, which leads to the very simple expression for this metric, $\mathfrak{D}_0$

$$\mathfrak{D}_0(R \parallel S) = \frac{1}{\mathbb{N}} \left| \sum_s log\left(\frac{R(s)}{S(s)}\right) \right|$$

(18)

where $\mathbb{N}$ is the number of values that the total set of variables can take on (consider it a vector.) This is always a metric since $\mathfrak{D}_P(R \parallel S) \neq 0$, <u>unless</u> $R(v) = S(v)$. An interesting class of metrics is generated by choosing a Gaussian reference.

If the functions R and S are also Gaussian we can illustrate a particularly simple expression for distances for the case of a single variable. Let the reference function be defined as a normal distribution with variance $\sigma^2$ and mean, $\mu$, designated

$$P(x) = \frac{1}{(2\pi)^{1/2}\sigma} exp\left\{-\frac{(x-\mu)^2}{2\sigma^2}\right\} \tag{19a}$$

and the functions to be measured are:

$$R(x) = \frac{1}{(2\pi)^{1/2}\sigma_1} exp\left\{-\frac{(x-\mu_1)^2}{2\sigma_1^2}\right\} \tag{19b}$$

$$S(x) = \frac{1}{(2\pi)^{1/2}\sigma_2} exp\left\{-\frac{(x-\mu_2)^2}{2\sigma_2^2}\right\} \tag{19c}$$

The distance between R and S then is: $\mathfrak{D}_G(R \parallel S) = \left|\int P(x) \left\{log\left(\frac{R(x)}{S(x)}\right)\right\} dx\right|$, which can easily be evaluated. Using the simple properties of Gaussians we have

$$\mathfrak{D}_G(R \parallel S) = \left|log\left(\frac{\sigma_2}{\sigma_1}\right) - \left\{\frac{\sigma^2}{2}\left(\frac{1}{\sigma_1^2} - \frac{1}{\sigma_2^2}\right) + \frac{(\mu-\mu_1)^2}{\sigma_1^2} - \frac{(\mu-\mu_2)^2}{\sigma_2^2}\right\}\right| \tag{20}$$

This explicit expression for distance has some simple special cases. First, if the *standard deviations* of R and S are *the same,* then the distance is dependent only on their mean values, independent of the standard deviation of the reference function. Likewise, if the *means* of R and S are *the same,* then the distance depends only on the standard deviation, independent of the mean of the reference function.

There is another, special case worth pointing out. If the reference function is chosen to be a Dirac delta function[2], which could be considered to be the limiting case of a Gaussian with vanishing standard deviation, the expression of Equation 20 simplifies further. The metric space is defined by the single parameter of the mean of the reference function, $\mu$. The distance expression, $\mathfrak{D}_\delta$, is then

---

[2] The key property of the Dirac delta function, $\delta(x-x_0)$, is that the integral over x with any function yields a specific value of the function, $\int \delta(x - x_0)f(x)dx = f(x_0)$.

$$\mathfrak{D}_\delta(R \parallel S) = \left|\left\{\frac{(\mu_1-\mu)^2}{\sigma_1^2} - \frac{(\mu_2-\mu)^2}{\sigma_2^2}\right\} - \log\left(\frac{\sigma_2}{\sigma_1}\right)\right| \qquad (21)$$

If the standard deviations of R and S are equal, the expression is extremely simple. To assure that this is a metric rather than a pseudometric we can chose the function space and the reference function such that $\mu \leqq \mu_1, \mu_2$, for example. In this case the distance between R and S is proportional to the squares of the distance from the reference mean. If the function space includes those with different $\sigma$ the ratios of $\mu$ to $\sigma$ defines the distance. We note that in this case there are many functions that are zero distance apart, but they are a very restricted class. If we set the reference mean at zero it is clear that the relevant measure of distance is just the squares of the ratios of the mean to standard deviation. This one-dimensional case has a simple geometric interpretation, which is illustrated in figure 1 in two different ways.

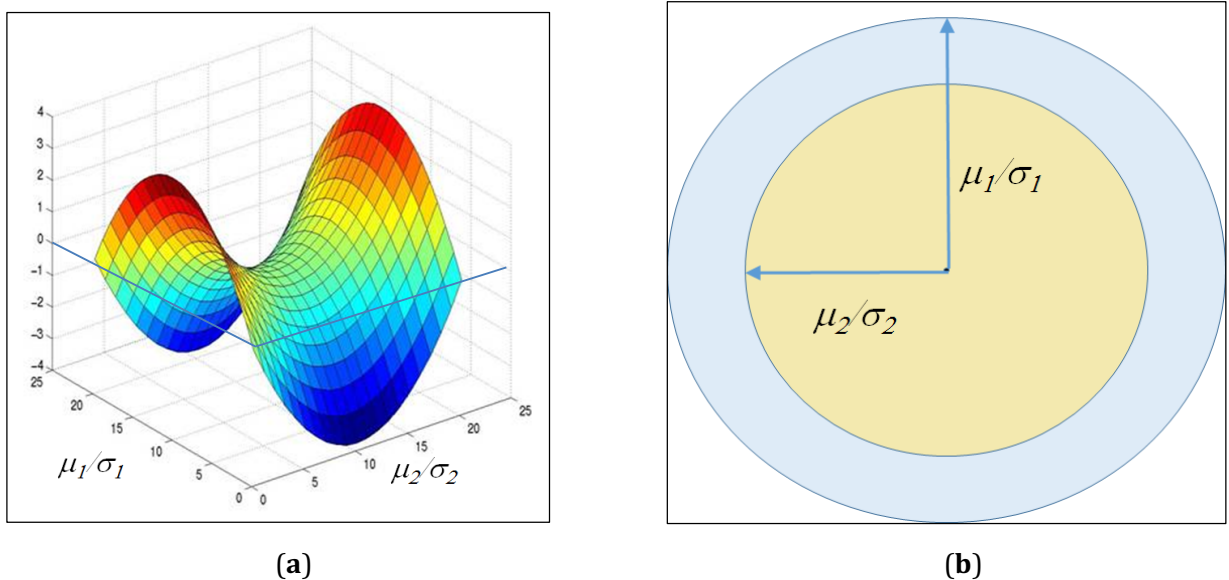

(a)　　　　　　　　　　　　　　　(b)

**Figure 1.** (a) The metric distance between Gaussian's R and S (Equation 21) for the Dirac delta function reference metric with mean at zero can be represented as a hyperbolic function (distance is the vertical axis where for simplicity the metric distance is the deviation from the zero plane – the absolute value) with a saddle point. (b) Another geometric metaphor for the distance is the area of the blue annular region divided by $\pi$ is the distance, where $\frac{\mu_1}{\sigma_1} > \frac{\mu_2}{\sigma_2}$. The expression for this area is simply: Area=$\frac{1}{\pi}\left(\frac{\mu_1}{\sigma_1}+\frac{\mu_2}{\sigma_2}\right)\left(\frac{\mu_1}{\sigma_1}-\frac{\mu_2}{\sigma_2}\right)$

Notice that with the exception of the single log term on the right-hand side of Equation 20, the expression is a quadratic form in the ratios of mean to standard deviation of R and S, and of the ratios of each of these standard deviations to the reference standard deviation.

In general, the Dirac delta reference function metric does not carry much information about the functions themselves, but if the function class is restricted it becomes both more interesting and useful. These logs, whose difference is the metric in Equation 16, are often called "surprisals" in information theory. So in this case, the metric is essentially how much more surprising is $R$ than $S$ at any specific point. We should mention that if multiple delta function metrics are used where the distance coordinates for each surprisal point $t$ distances between $R$ and $S$ leads naturally to a multi-dimensional space representation of the log ratios. A three-dimensional representation, for example, reflects the three chosen points where the functions are compared.

Another interesting metric space results from selecting all three functions, the reference and the measured functions, as Poisson distributions. These discrete valued functions, $P(k, \lambda) = \frac{\lambda^k}{k!} e^{-\lambda}$, yield a particularly simple metric distance. If the reference function has parameter $\lambda$, and the other two $\lambda_1$ and $\lambda_2$ the distance is simply

$$\mathfrak{D}_P(\lambda_1 \parallel \lambda_2) = |(\lambda_1 - \lambda \log \lambda_1) - (\lambda_2 - \lambda \log \lambda_2)| \tag{22}$$

Of course the distance vanishes when $\lambda_1$ goes to $\lambda_2$. If the reference $\lambda$, is much smaller than the other two, $\lambda \ll \lambda_1, \lambda_2$ the distance is linear in the difference between them, while if it is very much larger, $\lambda \gg \lambda_1, \lambda_2$ the distance is proportional to the difference of the logs of the $\lambda$'s. In these cases the distance is a true metric, with no distinct functions at zero distance. If the reference $\lambda$ were set to one, on the other hand, it is easy to see that there are pairs of functions, on either sides of one that have zero distance. For that choice of reference function then we have a pseudometric.

There are a very large number of possible special metrics based on a wide range of possible continuous distributions that could be used as reference functions, many of which lead to interesting functional expressions. To explore these further see the comprehensive list of such functions in the "Field Guide to Continuous Probability Distributions", which is available from Gavin Crooks' Website[3].

**Measuring the Independence of variable subsets.**

Next consider comparing the probability of a given, single variable, $R(X_1)$ with that of the conditional probability of that variable given the remaining set $v - \{X_1\}$ of $|v| - 1$ variables, $R(X_1|v - \{X_1\})$. If the chosen variable is independent of the others, so that we have $R(X_1) = R(X_1|v - \{X_1\})$, then the distance is zero: $\mathfrak{D}\big(R(X_1) \parallel R(X_1|v - \{X_1\})\big) = 0$. This result is independent of the reference function, $P(v)$.

Also, we have

$$\mathfrak{D}\big(R(X_1) \parallel R(X_1|v - \{X_1\})\big) = \left| \sum_s P(s) \left\{ \log \left( \frac{R(s_1)}{R(s_1|s - \{s_1\})} \right) \right\} \right|$$

$$= \left| \sum_s P(s) \left\{ \log \left( \frac{R(s_1)R(s - \{s_1\})}{R(s_1|s - \{s_1\})R(s - \{s_1\})} \right) \right\} \right|$$

where $s_1$ denotes a state of a single variable $X_1$ and $s - \{s_1\}$ denotes the state of all other variables of $v$.

Therefore, $\mathfrak{D}\big(R(X_1) \parallel R(X_1|v - \{X_1\})\big) = \mathfrak{D}\big(R(X_1)R(v - \{X_1\}) \parallel R(v)\big)$. Generalizing from a single variable to a subset $v'$ we have:

$$\mathfrak{D}\big(R(v') \parallel R(v'|v - v')\big) = \mathfrak{D}\big(R(v')R(v - v') \parallel R(v)\big) \tag{23}$$

which is, of course dependent on the reference function, $P$, except in the limit where the distance goes to zero.

---

[3] This compendium can be found on the website http://threeplusone.com/FieldGuide.pdf

**Comparing approximations from different truncated series.**

An application of these metric spaces lies in the area of statistical physics, for example, that considers reduced probability distribution functions to approximate the true distribution functions. Both high degree of interactions, highly multivariable, and non-equilibrium problems defined by trajectories, could be directly approached with this apparatus. These approximations have often involved physically motivated simplifying truncation relationships, like those discussed above. The formalism developed here can be used to calculate the distance between probability functions that are truncated at different levels of approximation. This allows an assessment of the convergence of higher level truncations, in terms of the distance converging to zero.

The approximate functions that result from truncations of the variable number expansion at different numbers of variables can now be directly compared with a quantitative metric. The truncations described above define the forms of density functions as factors. The actual pdf's are determined by the true, or reference pdf.

Comparing distributions truncated at the first and second order, the probability functions $P'_1$ and $P'_2$ are determined by the factorizations of Equations 6c and 7a. The distance between these two truncation approximations, relative to the reference function, then is:

$$\mathfrak{D}(P'_2 \parallel P'_1) = |D(P'_2 \parallel P) - D(P'_1 \parallel P)| = |A_2 - A_1| \tag{24a}$$

Referring to Equations 6a and 8c this expression simplifies to

$$\mathfrak{D}(P'_2 \parallel P'_1) = |A_2 - A_1| = \sum_{i>j}\left(H'(X_i) + H'(X_j) - H'(X_i, X_j)\right) = \sum_{i>j} I'(X_i, X_j)$$

$$\mathfrak{D}(P'_2 \parallel P'_1) = \sum_{\tau_2 \subset \nu} I'(\tau_2) \tag{24b}$$

Recall that this is the sum of "cross" mutual information between all pairs of variables defined by the reference function. The distance of Equation 24b represents the distance between functions of pairwise dependence and independence. In general, the distance between two different truncation approximations can be seen easily from the expansion of Equation 4. The distance is simply the absolute value of the sum of the terms present in only one of the truncated series.

**Application to Networks**

We can use our metric spaces to devise a simple and direct way of estimating the distance between two networks, which is a problem that has attracted attention for many years. We begin by considering networks in terms of subsets of dependent variables, where the variables are nodes, so that if there are only pairwise dependencies we have a graph. Furthermore we can consider the measures of dependence as weights for the edges, so that the mutual information between variable pairs provides these weights. For higher-degree dependencies the corresponding network is a hypergraph. Let us consider here how our metric spaces apply to graphs. The general formalism described here can be used for hypergraphs by direct analogy. Extending the analysis to hypergraphs adds some additional consideration that we do not address here, but the parallel is clear. A graph describing the dependencies present in a dataset, for example, would result from truncation of the divergence at the *m*=2 level. Following our previous discussion, the divergence can be used then to quantitate the approximation represented by the graph. Moreover, the metrics we have defined now present a simple way to calculate real metric distances between graphs.

Let us consider an example: let us choose for simplicity a uniform probability density over the *n* variable set, $v$, as a reference function, resulting in a very simple expression for a metric, $\mathfrak{D}_0$, as shown in equation 18. If $\mathbb{N}$ is the number of values that the set of variables can take (consider it a vector) and we have two acyclic graph, *R* and *S*, defined by density functions, as in equation 8a:

$$R(X_1, X_2, X_3 ... X_n) = \prod_{i>j} R(X_i | X_j) \qquad S(X_1, X_2, X_3 ... X_n) = \prod_{i>j} S(X_i | X_j) \qquad (25)$$

Using the uniform density metric, $\mathfrak{D}_0$, and the relationship between the mutual informations for these distributions, we have then

$$\mathfrak{D}_0(R \parallel S) = \frac{1}{\mathbb{N}} \sum_s \left| \sum_{i>j} \log \frac{R(X_i|X_j)}{S(X_i|X_j)} \right| = \frac{1}{\mathbb{N}} \left| \sum_{i>j} I_R(X_i, X_j) - I_S(X_i, X_j) \right|$$

(26)

where the sums are over the weights of the edges. Note that using graphs with the weights as mutual information between nodes to describe a dataset is exactly like the Chow-Liu approach. In our example the distance is simply a sum of differences for the same sets of nodes in the two graphs. This is a simple result, but with an interesting subtlety. These differences may be positive or negative and the $\mathfrak{D}_0$ is the absolute value of the overall sum. The absence of an edge (zero mutual information) then in one graph may be compensated for by a different absence in the other, to leave the distance the same. Our formalism guarantees that the distance is a true metric distance. Other reference functions, which also produce metrics, lead to more complex results that are not so easy to visualize, and the extension to hypergraphs, which is a natural extension of the above, leads to results that are increasingly difficult to see. In any case, this demonstrates the use of our formalism in network comparison based on information functions. The further applications of these network comparison results will be explored in a later paper.

**Conclusions**

The Kullback-Leibler divergence as a means of comparing probability density functions has played a central role information theory and been used for several practical purposes in data analysis, machine learning, and model inference. It has provided ways to explore some key ideas in fields from information theory to thermodynamics. We show here that it can continue to yield new results. The divergence can be expanded in the number of interacting variables, yielding a systematic hierarchy of truncations, approximations to the probability density function, which is effectively a hierarchy of factorizations. Factorizations, since they focus on the kinds and degrees of independence are central and can be thought of as hierarchies of spaces of ever more complex functions, with evermore complex dependencies. The relationship between the set of entropies and the set of interaction informations through the Möbius inversion relation is a fundamental symmetry that is manifest here, but the full symmetry spectrum is deeper yet. It reflects a number of relationships with other information-related measures, that are based on this symmetry [9,10]. Since these relations can also be used to express the cross entropy differently, they should generate different expansions of the divergence, with different structures. This intriguing area may itself yield additional, useful applications, and has yet to be explored.

As we noted in the introduction there are several areas of potential application of these ideas. The relation of one level of the truncation hierarchy to the Chow-Liu approximation [15] we noted immediately suggests the extension of the Chow-Liu algorithm to higher levels – Chow-Liu-like hypergraphs. This remains to be explored fully and will be addressed in a future publication. Other applications to networks and network inference are suggested by the notion of the metric classes based on specific reference functions. It is interesting that the divergence provides the basis for a new finite difference metric that gives measures of distances between pdf's, real or estimated, continuous or discrete. We give an example of the application of these metrics to a graph comparison problem. The example shows that the entire formalism can be brought to bear on a wide range of network problems. It should be possible to simplify graph distance measures, given a set of specific constraints, by optimizing the choice of a reference function. Tailoring the metric to specific classes of graphs, for example, should enable simplification of model inference in some cases. These ideas will be addressed in future work.

If we added the constraint of specific function forms for the pdf's - the exponential family of functions like Gaussians, for example - the natural extension leads to a number of specific approximations and metric form. The considerations here raise the question of the strategy that should be used to select a reference function. There are at least two considerations. If it is important that a true metric, rather than a pseudometric, be provided then the reference function and the function space should be selected to provide that property. It could be as simple as picking the right parameter range for the pdf, as we illustrated for the Poisson and Gaussian pdfs. Another consideration is the accuracy of the calculated distances. This will depend on what the relevant functions are expected to be, so that a reference function might be chosen near this region in function space to avoid having to subtract two large divergences to find the distance. These issues are important and practical, and somewhat problem specific, for the effective use of these concepts. They will be systematically considered in future work.

The relationship of our metric (Eqn. 18) to the Fisher information metric can be obtained from the convergence to zero of this distance. There are a wide variety of metrics that can be derived by symmetrizing the divergence in various ways. The Jensen-Shannon divergence is one of these, but there are several others that use variations on the theme of averaging the cross entropy terms in various ways. The proposed metric here is the first to our knowledge to use a third probability density function to define the character of the metric space, though there has been some consideration of the information-geometric interpretation of the difference between K-L divergences [18]. The connection of K-L divergence differences using a third distribution to expected log-likelihood ratios and to their use in building Riemannian metrics has also been discussed previously [17,18]. These approaches are quite distinct from ours but may be connected by future work.

In general our approach has the advantage, as indicated by the simple examples shown here, that the metric space can be tailored to the character of any function space. We suggest that Equation 18 defines what could be interpreted in a general sense as a finite difference form of the Fisher metric. The metric can also be used directly to compare Bayesian estimators as the pdf is iteratively updated, to measure convergence.

The application of the general approach described here to a wide range of multivariable problems, including data analysis, model inference, multivariable physical problems, and problems involving complex biological systems, should be useful in providing new analysis methods and new insights.

**Acknowledgements**: This work was supported in part by the NIH Common Fund, the Extracellular RNA Communication Consortium (ERCC) 1U01HL126496-01, the Bill and Melinda Gates Foundation, and the Pacific Northwest Research Institute. The authors thank an anonymous referee for a number of specific corrections and suggestions.

**Conflicts of Interest**: The authors declare no conflict of interest. The founding sponsors had no role in the design of the study; in the collection, analyses, or interpretation of data; in the writing of the manuscript, and in the decision to publish the results.